\documentclass[conference,9pt]{IEEEtran}
\IEEEoverridecommandlockouts

\usepackage{booktabs} 
\usepackage{cite}
\usepackage{amsmath,amssymb,amsfonts}
\usepackage{algorithmic}
\usepackage{graphicx}
\usepackage{textcomp}
\usepackage{xcolor}
\def\BibTeX{{\rm B\kern-.05em{\sc i\kern-.025em b}\kern-.08em
    T\kern-.1667em\lower.7ex\hbox{E}\kern-.125emX}}
\begin{document}

\title{Typical vs. Atypical Disfluency Classification: Introducing the
 IIITH-TISA Corpus and Temporal Context-Based Feature
 Representations\\
}
\author{
\textit{Priyanka Kommagouni$^1$, Vamshiraghusimha Narasinga$^1$, Purva Barche$^1$, Sai Akarsh C$^1$, Anil Vuppala$^1$}\\
$^1$LTRC, International Institute of Information Technology - Hyderabad\\
\{parvathipriyanka.b,narasinga.vamshi,purva.sharma, sai.akarsh\}@research.iiit.ac.in\\
\ anil.vuppala@iiit.ac.in}

\maketitle

\begin{abstract}
Speech disfluencies in spontaneous communication can be
 categorized as either typical or atypical. Typical disfluencies, such as
 hesitations and repetitions, are natural occurrences in everyday speech,
 while atypical disfluencies are indicative of pathological disorders like
 stuttering. Distinguishing between these categories is crucial for improv
ing voice assistants (VAs) for Persons Who Stutter (PWS), who often
 face premature cutoffs due to misidentification of speech termination.
 Accurate classification also aids in detecting stuttering early in chil
dren, preventing misdiagnosis as language development disfluency. This
 research introduces the IIITH-TISA dataset, the first Indian English
 stammer corpus, capturing atypical disfluencies. Additionally, we extend
 the IIITH-IED dataset with detailed annotations for typical disfluencies.
 We propose Perceptually Enhanced Zero-Time Windowed Cepstral Co
efficients (PE-ZTWCC) combined with Shifted Delta Cepstra (SDC) as
 input features to a shallow Time Delay Neural Network (TDNN) classifier,
 capturing both local and wider temporal contexts. Our method achieves
 an average F1 score of 85.01\% for disfluency classification, outperforming
 traditional features.

\end{abstract}


\section{Introduction}

Human communication is a complex interplay of linguistic and cognitive processes, influenced by vocal nuances and environmental contexts. Speakers often experience disfluencies—pauses, revisions, and repetitions—as they navigate real-time language production. Research indicates that these brief disruptions occur in approximately 6\% of fluent words during typical speech, reflecting the cognitive demands of everyday conversation\cite{lickley2017disfluency}. Atypical speech, such as stuttering, presents more severe disruptions and affects around 70 million people, constituting 1\% of the global population \cite{yairi2013epidemiology}. Stuttering disfluencies differ significantly from typical ones in physical and phonetic characteristics along with their underlying causes. The stark contrast between them underscores the importance of understanding and differentiating these patterns in both clinical and technological contexts.
Research on improving voice assistants for people who stutter has largely focused on detecting and correcting disfluencies. However, through consultations with The Indian Stammering Association (TISA), we discovered a more pressing issue: many individuals who stammer face considerable frustration due to premature interruptions by the endpointer—the system component responsible for determining when a user has finished speaking. This insight highlights the need for more inclusive voice assistant design that accounts for diverse speech patterns. Recently, \cite{apple} has attempted to address
 premature cutoffs by fine-tuning interruption thresholds. While this approach may improve the experience for some individuals who stutter, it may not be universally effective across all stuttering severities and could potentially slow response times for non-stutterers. To address this limitation, one possible solution could be incorporating
 a classifier after disfluency detection, allowing for separate threshold adjustments for stutterers and non-stutterers.

Understanding the disparities between these forms is also crucial for early detection of stuttering in children. Stutter is often misdiagnosed as language development disfluency amongst children, leading to delayed intervention. Research indicates that children as young as two years old may begin to demonstrate an awareness of their stuttering condition\cite{boey2009awareness}, potentially leading to communication withdrawal. This can complicate stuttering diagnosis, which relies heavily on speech pattern analysis.

While \cite{lickley2017disfluency} focused on observable features and manifestations of the typical and atypical disfluencies, giving a diagnostic perspective to stuttering,\cite{starkweather1982stuttering} discusses the etiology of stuttering that emphasises the changes in laryngeal structure. In a comparative analysis of disfluency in stutterers and non-stutterers, \cite{kuniszyk1996comparison} suggest the area under the speech envelope, together with total phonation time and pause durations for disfluency classification. In \cite{wingate2002foundations}, authors proposed an average of approximately 10 stuttering events per 100 words as a reasonable estimate derived from various studies. While these values exhibit distinguishing features for the current classification, they lack robustness due to significant impact from speaker-to-speaker variability in frequency and statistical measures\cite{shriberg2005spontaneous}.
To the best of the authors' knowledge, no published research has employed machine learning or signal processing techniques to address this specific classification problem. 

The primary challenge, common in pathological domain, is the scarcity of datasets tailored for this particular classification task. To minimize inter-dataset variability, it's crucial to have both typical and atypical disfluencies from the same linguistic context. Current literature shows a notable bias towards stuttering studies in British and American English contexts, largely due to the availability of datasets \cite{lea2021sep}, \cite{9528931}. However, India boasts one of the world's largest English-speaking populations, with approximately 83 million individuals utilizing English as their second language \cite{mathews2018language}. This demographic reality emphasizes the increasing importance and relevance of profiling disfluencies within the Indian English linguistic context. We introduce IIITH-TISA, the first Indian English stammer dataset, created in collaboration with The Indian Stammering Association (TISA). This novel dataset has been meticulously annotated to include five types of disfluencies: Filled pauses, Prolongations, Part-Word Repetitions, Phrase Repetitions, and Word Repetitions along with segments with no disfluencies, from a total of 10 hours of speech data from Persons Who Stutter (PWS). It also includes annotations for segments with no disfluencies, providing a comprehensive representation of stuttered speech patterns in the Indian English contexts. We also significantly enriched IIITH-IED dataset, which captures typical disfluencies curated from freely available lectures from the Government of India’s NPTEL initiative\cite{garg2021towards}. Our datasets' curation and annotation guidelines align with the SEP28k corpus, the largest public resource for stuttered speech\cite{lea2021sep}. This alignment extends our datasets' utility beyond this study. By adhering to established standards, we enhance the datasets' interoperability and potential for comparative studies. Our handcrafted disfluency embeddings, which leverage both short and long temporal contexts of Perceptually Enhanced Zero-Time Windowed Cepstral Coefficients (PE-ZTWCC), outperformed traditional state-of-the-art features in terms of F1 scores.

Our primary contributions are:
\begin{itemize}
\item IIITH-TISA: Collection, annotation, and curation of Indian English stammer dataset.
\item Extended IIITH-IED: An expanded dataset of spontaneous non-pathological disfluencies with increased key events.
 \item PE-ZTWCC: Novel feature set perceptually enhancing zero time windowing cepstral coefficients to mimic human auditory
\item SDC-TDNN: Insights into Leveraging short and long temporal contexts by optimizing context-dependent parameters of both SDC and TDNN.

\end{itemize}
\section{IIITH-TISA CORPUS}
The Indian Stammering Association (TISA), established in 2008,
 is an organization dedicated to raising awareness about stammering,
 offering resources and peer support, and advocating for the rights
 of individuals who stutter across India. Through collaboration with
 this organisation, we gained permission to access their TISA Online
 Program Group (TOPG) 2.0 video series. These videos, covering
 topics such as self-presentation, job interview practice, and public
 speaking, were uploaded by the PWS members of the association to
 be reviewed for technical feedback. We then developed a rigorous
 pipeline to transform these recordings into a research dataset. This
 process involved carefully screening videos for minimal background
 noise and absence of code-switching, while excluding content fea
turing stammerers under current therapy or therapy-related material.
 The selected YouTube videos were downloaded as audio files. All
 the audio files were converted to WAV format and their audio quality
 was ensured. They were next resampled to 16kHz, converted to
 single channel, and anonymized by removing personal information.
 This meticulous selection yielded 10 hours of data from 30 unique
 Persons Who Stutter (PWS). Using Audacity software, we annotated
 disfluency events by generating label files with time stamps for each
 event within the audio files. Adhering to the SEP-28k protocol,
 we standardized clips to 3 seconds, carefully trimming each file to
 include the disfluency event based on the label file time stamps.
 To ensure natural speech segments, we employed voice activity
 detection (VAD) for trimming at appropriate endpoints. Shorter files
 were padded with silence for consistency, and any chopping-induced
 transients were removed. The resulting IIITH-TISA dataset comprises
 3,251 audio clips, each 3 seconds long and categorized into one of
 the disfluency event types, representing a significant contribution to
 stammering research in the Indian English context.

 \subsection{Annotation}
Drawing inspiration from \cite{bayerl_KSoFKasselState_2022} towards annotation of our dataset, we implemented a comprehensive annotation process to identify and categorize disfluency events in the audio recordings. Our approach involved marking not just the occurrence of each disfluency, but also precisely documenting its duration by recording the exact start and end times within each audio file. To carry out this task, we assembled a diverse team of six undergraduate students, including those with and without speech-related academic backgrounds.

 Prior to beginning the annotation work, all team members underwent an intensive three-hour training session. This training covered two key areas: technical proficiency with the Audacity software tool, and perceptual recognition of various disfluency types. To ensure accuracy, we closely monitored the initial annotations, collaborating with a speech-language pathologist to review the results. In cases where ambiguities or inconsistencies were identified, we provided additional guidance and retraining to the annotators, focusing on clarifying any misunderstandings.

 Our annotation process was designed to be iterative, incorporating regular inter-annotator checks. This approach allowed us to verify and maintain consistency in how different types of disfluencies were identified and labelled across the dataset. Through this rigorous methodology, we aimed to create a high-quality, reliably annotated corpus for disfluency research.

During our analysis of the IIITH-IED dataset, which serves as our primary source for typical disfluency patterns in this study, we noticed a lower frequency of certain disfluency types. Specifically, we observed fewer instances of Prolongation and all three categories of Repetition disfluencies. To address this imbalance and enhance the dataset's comprehensiveness, we expanded our annotation efforts to include 30 additional speakers. We refer to this expanded version as IIITH-IED-E. 
TABLE \ref{tab:disfluency_counts} presents the statistics about the
 number of disfluency events in each of the datasets.  
Both datasets will be made accessible to other researchers upon request.



\begin{table}[ht]
\caption{Disfluency Event Counts Across Three Datasets}
\label{tab:disfluency_counts}
\centering

\resizebox{\linewidth}{!}{
\begin{tabular}{cccc}
\hline
\textbf{Disfluency Type} & \textbf{IIITH-TISA} & \textbf{IIITH-IED} & \textbf{IIITH-IED-E} \\ \hline
\textbf{Filled Pauses} & 626 & 1277 & 1277 \\ 
\textbf{Prolongations} & 313 & 71 & 293 \\ 
\textbf{Part-Word Repetitions} & 818 & 164 & 358 \\ 
\textbf{Phrase Repetitions} & 197 & 76 & 258 \\ 
\textbf{Word Repetitions} & 96 & 211 & 618 \\ 
\textbf{No Disfluencies} & 1201 & 2013 & 2013 \\
\hline
\end{tabular}
}
\end{table}

\section{PROPOSED FEATURE EXTRACTION}
Stutterers showcase notable laryngeal tension, evidenced by pitch changes, aperiodic sounds, and altered voice quality\cite{starkweather1982stuttering}. They also exhibit greater prosodic irregularities and over-vigilant self-monitoring, leading to prosodic breaks and degraded phonetic cues before the actual disfluencies \cite{russell2005magnitude,arbisi2005comparison}.
Drawing insights from literature, we propose a novel Perceptually Enhanced Zero-Time Windowing Cepstral Coefficients (PE-ZTWCC) combined with shifted delta cepstra (PE-ZTWCC+SDC) as our distinguishing features for both the datasets. ZTW's high temporal resolution and group delay spectral estimation can effectively capture vocal tract excitation effects and spectral envelope decay in tense voice\cite{kadiri2018breathy,simha2023enhancing}. Stutterers' disfluencies disrupt phrasing and tonal context more variably than those of non-stutterers, leading to greater interference with listeners' comprehension of information structure \cite{arbisi2005comparison}. This insight was reinforced when we observed that Perceptual Linear Prediction features (PLP) outperformed Mel-Frequency Cepstral Coefficients (MFCCs) in our initial analysis. Taking this as a cue, we applied perceptual enhancement to the Zero-Time Windowing (ZTW) spectrum. Our experiments demonstrate that PE-ZTWCC improve F1 scores by emulating the non-linear frequency scaling of the human auditory system.

\subsection{ PE-ZTWCC Feature Extraction}
The proposed PE-ZTWCC features are extracted from a speech signal in two steps, ZTW spectrum is first extracted and then perceptual enhancement on the spectrum is carried out. It begins with pre-emphasis to remove low-frequency trends from the speech signal. At each instant, a speech segment of the duration 
$L$ ms is considered. Signal $s[n]$ is defined for $n = 0, 1, . . . , M-1$, where $M = L \times \frac{fs}{1000}$, with $fs$ being the sampling frequency. Then, a heavily decaying window $w_{1}[n]$ is applied to a speech segment\cite{bayya2013spectro}. 
\begin{equation}
w_1[n] = \begin{cases}
0, for \quad n = 0 \\
\frac{1}{4 \sin^2\left(\frac{\pi n}{2N}\right)}, for \quad n = 1, \ldots, N-1
\end{cases}
\end{equation}
A discrete Fourier transform (DFT) of a signal $s[n]$ with $N$ samples, where $N \gg M$.
Multiplying $s[n]$ by a window $w_{1}^{2}[n]$ is equivalent to four times integration in the frequency domain. However, truncating the signal at the instant $n = M-1$ may result in a ripple effect in the frequency domain. To mitigate ripple effect caused by truncating the signal at the instant $n = M-1$ , we use a window $w_2[n]$, the square of a half-cosine window which improves the spectral estimation of the signal.
\begin{equation}
w_2 [n] = 4 \cos^2 \left( \frac{\pi n} {2 M}\right) where \quad n = 0, 1, \ldots, M-1.
\end{equation}
The spectrum is estimated using the numerator of group delay (NGD) function, and double differentiation of NGD highlights formant peaks. Finally, the Hilbert envelope of double-differenced NGD yields the ZTW spectrum, denoted as $X[n,k]$. 
The Mel frequency warping is carried out to achieve the non-linear frequency scaling property of the human auditory system \cite{makur2001warped}. The warped spectrum is given by 
\begin{equation}
X_W[n,k]   = \psi_m {X[n,k]} 
\end{equation}
Where m is a Mel warping operator.
An equal-loudness pre-emphasis contour is used to model the non-uniform
 sensitivity of human hearing at different frequencies\cite{hermansky1990perceptual}. 
\begin{equation}
X_{WE}[n,k]   = \psi_e  {X[n,k]}
\end{equation}
The power law non-linearity with
exponent 1/5 when operated on  $X_{WE}[n,k]$. simulates the non-linear relationship between sound intensity and perceived loudness\cite{kim2016power}. This results
in 
\begin{equation}
X_{WEP}[n,k]=X_{WE}[n,k]^{\frac{1}{5}}
\end{equation}

The inverse Fourier transform of the logarithmic power spectrum is used to obtain the cepstral representation. We reduced the feature dimension of instantaneous cepstral representations by subsampling\cite{alluri2017detection}. A 13-dimensional cepstral feature is derived through a liftering operation, represented as:
\begin{equation}
c[n,k] = IFFT(log(X_{WEP}[n,k]))
\end{equation}

\section{Classification Framework}
\begin{figure}[h!]
    \centering
    \includegraphics[width=\linewidth]{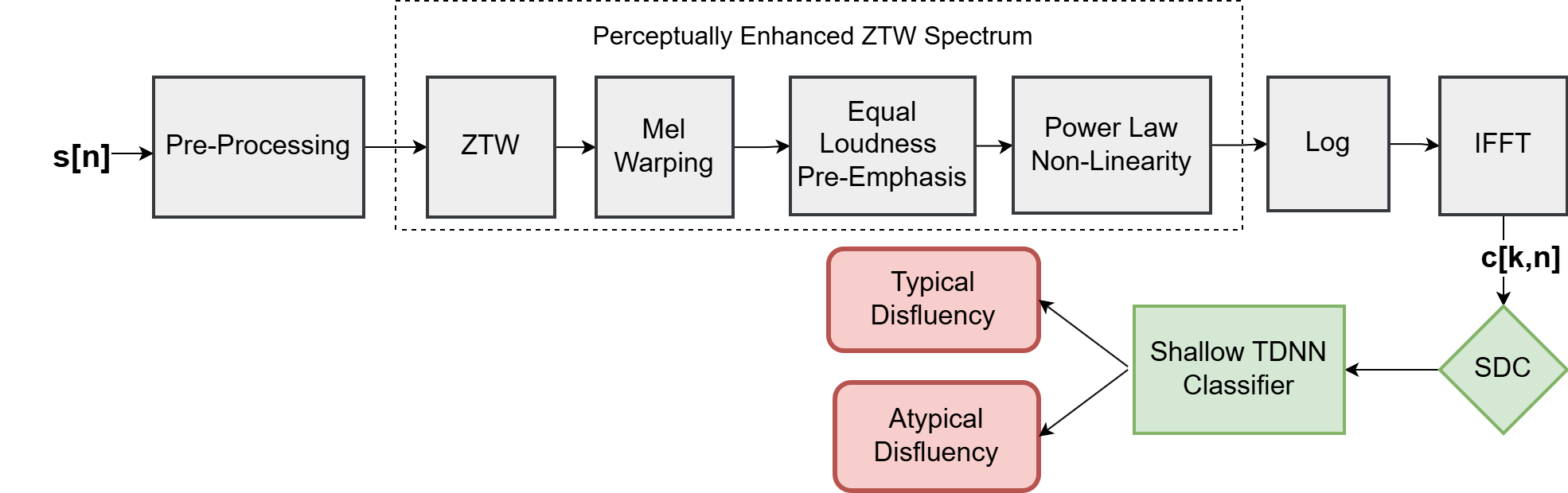}
    \caption{Block Diagram of the Proposed Architecture}
    \label{fig:a}
\end{figure}

For our classification task, capturing temporal dependencies is crucial to understand the frame-to-frame tonal variations surrounding both typical and atypical disfluencies.The limited size of our dataset increases overfitting risk, particularly with complex neural networks. To address this, we propose a solution that leverages Shifted Delta Cepstra (SDC) features in conjunction with a shallow Time Delay Neural Network (TDNN) architecture.TDNNs already proved to work better with limited data by using a modular and incremental design from sub-components. Computations in TDNN are reduced by selection of appropriate timesteps at which activations are computed\cite{peddinti2015time}. Shifted Delta Cepstra (SDC) features can compensate for fewer layers in a shallow Time Delay Neural Network (TDNN) by providing richer local temporal context at the input level. By meticulously adjusting the N-d-p-k parameters of SDC, alongside kernel size and dilation rate of TDNN we can capture complex temporal patterns typically associated with deeper networks while mitigating the risk of overfitting and maintaining computational efficiency.The block diagram describing the steps involved in proposed architecture is shown in Fig.\ref{fig:a}.
\subsection{ Shifted Delta Cepstra }

Research has demonstrated that shifted delta cepstra (SDC) effectively capture temporal variations in speech across multiple frames\cite{torres2002approaches}. In this study, SDC features are derived from the PE-ZTWCCs. The computation of these features is governed by four parameters, referred to as \(N\)-\(d\)-\(p\)-\(K\). Specifically, at a given time \(t\), delta computations are performed between the cepstral coefficients of the \((t + ip + d)\)-th frame and the \((t + ip - d)\)-th frame. By varying \(i\) from 0 to \(K-1\) and stacking the resulting values, the delta coefficients are obtained.

The SDC features, denoted as \(c(t, i)\), are calculated using the formula:
\begin{equation}
c(t, i) = C(t + ip + d) - C(t + ip - d)
\end{equation}

where \(N\) represents the dimension of the static cepstral coefficients; \(d\) is the delay or advance relative to the current frame; \(p\) is the shift between consecutive delta computations; and \(K\) denotes the number of such delta computations concatenated to form \(N \times K\) dimensional SDC coefficients. This study initially employs a 13-1-3-7 configuration, resulting in 91 SDC features. Combining the 13-dimensional static PE-ZTWCCs with these 91-dimensional SDCs yields a 104-dimensional feature vector for each frame. Then explores the impact of varying \(d\) and \(K\) values on the temporal context captured by the features by keeping \(N\) and \(p\) values constant.

\subsection{ Time Delay Neural Network }
SDC features capture wider temporal dynamics than traditional frame-level features, effectively pre-processing some of the temporal relationships that a deeper TDNN would typically learn. This allows a shallower TDNN to focus on extracting higher-level patterns rather than basic temporal correlations. After extensive experimentation with various one-dimensional convolutional layers, exploring different kernel sizes and dilation rates, we identified the following architecture as the most effective for our classification task. The initial layer employs 64 filters (kernel size 5, dilation rate 2) to capture a broad temporal context, while the second layer expands this with 128 filters (kernel size 7, dilation rate 3) to detect more complex patterns. To enhance model stability and prevent overfitting, we incorporate batch normalization, max pooling, and dropout techniques. Additionally, L2 regularization is applied to both convolutional and fully connected layers, further mitigating overfitting risks by penalizing large weights in the loss function. The flattened outputs from convolutional layers are processed through two fully connected layers (128 and 64 units respectively) with ReLU activation and dropout. The model concludes with a single-unit output layer using sigmoid activation for classification. We assessed the model's performance using both accuracy and F1 score metrics to ensure a comprehensive evaluation of its effectiveness in disfluency classification.


\subsection{ Baseline Features for Comparison}

 We evaluated our proposed features against traditional ones such
 as MFCC, PLPCC along with SFCC, another established spectro-temporal feature set. These features were assessed both in their
 standard form and When augmented with Shifted Delta Cepstra
 (SDC).We investigated various temporal combinations by adjusting
 the eN-d-p-K values of SDC and explored different kernel sizes and
 dilation rates in TDNN to determine their impact on classification
 performance across these feature sets.


\section{Results and discussions}
Our experiments compared the performance of various feature extraction techniques in classifying typical and atypical disfluencies across three categories: Repetition(combination of word, part-word and phrase repetitions), Filled-Pause, and Prolongation. Amongst these three events, results of Repetitions were better than the other two. This could be due to the prominent difference between typical and atypical repetititons. While typical repetitions in English often involve pauses and restarts of unstressed function words\cite{shriberg1994preliminaries}, Atypical repetitions in stutterers reflect difficulties in vocal fold transition, often leading to preemptive laryngeal closure and repetitive sounds\cite{freeman1979phonation}.


\begin{table}[ht]
\centering
\caption{F1 Scores (in\%) for different features with TDNN Classifier}
\label{tab:f1scores}
\begin{tabular}{cccc}
\hline
\textbf{Features} & \textbf{Repetition} & \textbf{Filled-pause} & \textbf{Prolongation} \\

\hline
MFCC & 77.34 & 73.32 & 72.24 \\
PLP Features & 80.12 & 80.36 & 78.74 \\

ZTWCC & 81.22 & 80.85 & 79.20 \\
PE-SFCC& 76.09 & 73.33  & 74.21 \\
PE-ZTWCC& \textbf{83.22} & \textbf{81.23} & \textbf{81.15} \\

\hline
\end{tabular}
\end{table}
The results, presented in TABLE \ref{tab:f1scores}, demonstrate a clear hierarchy in feature effectiveness. Perceptually Enhanced Zero-Time Windowed Cepstral Coefficients (PE-ZTWCC) consistently outperformed other features, achieving the highest F1 scores across all disfluency types.This superior performance can be attributed to PE-ZTWCC's ability to capture both spectral and temporal characteristics of speech while mimicking human auditory perception. 
\begin{table}[ht]
\centering
\caption{F1 Scores (in\%) for SDC combined Features}
\label{tab:f1scores2}
\begin{tabular}{cccc}
\hline
\textbf{Features} & \textbf{Repetition} & \textbf{Filled-pause} & \textbf{Prolongation} \\
\hline
MFCC+SDC & 80.34 & 80.60 & 77.23 \\
PLP+SDC & 83.66 & 84.43 & 81.80 \\
ZTWCC+SDC & 85.32 & 82.76 & 80.89 \\
PE-SFCC+SDC& 78.09 & 73.07  & 74.21 \\
PE-ZTWCC+SDC & \textbf{86.98} & \textbf{85.45} & \textbf{82.78} \\
\hline
\end{tabular}
\end{table}

Building upon our initial findings, we further investigated the impact of incorporating Shifted Delta Cepstra (SDC) with our feature sets (TABLE \ref{tab:f1scores2}). The addition of SDC consistently improved classification performance across all feature types and disfluency categories. PE-ZTWCC+SDC emerged as the top performer, with F1 scores of 86.98\%, 85.45\%, and 82.78\% for repetition, filled-pause, and prolongation, respectively. It's worth noting that ZTWCC outperformed SFCC in stutter detection, aligning with findings from previous research \cite{simha2023enhancing}, further suggesting that the choice of base feature remains crucial even when applying temporal context enhancements. These results accentuate the potential of PE-ZTWCC+SDC as a powerful tool for typical atypical disfluency classification.
To assess the TDNN's effectiveness, we tested Deep Neural Networks (DNN) and Bidirectional Long Short-Term Memory (BiLSTM) as alternative classifiers. These models overfitted our data due to high computational complexity and the limited size of our dataset. Consequently, their performance fell short of the TDNN's results, emphasising the TDNN's robustness in handling our specific data constraints.


\begin{table}[ht]
\centering
\caption{F1 Score (IN \%) for N-d-p-K combinations on PE-ZTWCC+SDC features with TDNN Classifier}
\label{tab:SDC}
\begin{tabular}{cccc}
\hline
\textbf{SDC Combinations} & \textbf{Repetitions} & \textbf{Filled Pause} & \textbf{Prolongation} \\
\hline
13-\textbf{1}-3-7 & 80.89  & 81.28 & 79.87 \\
13-\textbf{2}-3-7 & 81.34 & 81.89 & \textbf{82.78} \\
13-\textbf{3}-3-7 & 79.12 & 80.12 & 79.23 \\
\hline
13-2-3-\textbf{5} & 82.56 & \textbf{85.45} & 80.17\\
13-2-3-\textbf{6} & \textbf{86.98} & 84.67 & 80.34 \\
13-2-3-\textbf{7}   & 81.34 & 81.89 & \textbf{82.78} \\
\hline
\end{tabular}
\end{table}
We investigated the configurations of SDC parameters on PE-ZTWCC features keeping TDNN as classifier. Two important parameters d which indicates number of delays from current frame and K which are number of delta computations per concatenation were varied in steps and top results are tabulated in Table \ref{tab:SDC}. N, static cepstral coefficients are maintained to be 13 and shift between delta computations, p, is fixed to 3. Among the combinations, 13-2-3-6 achieves the highest F1 score for Repetitions with 86.98\%. For Filled Pause, 13-2-3-5 yields the best result at 85.45\%, while 13-2-3-7 excels in Prolongation with a top score of 82.78\%. Atypical prolongations cause pitch accents and rushed subsequent speech, contrasting with typical prolongations used for message planning \cite{arbisi2005comparison}. As prolongation usually lasts longer, it benefits from a longer context. Overall, these results illustrate the impact of varying the N-d-p-K parameters on the classification performance for different types of disfluencies, highlighting the combination 13-2-3-6 as particularly effective across multiple categories.
\section{Conclusion}
This research introduces two key datasets: the IIITH-TISA, India's first Indian English Stammer Corpus, and the extended IIITH-IED for typical disfluencies. By leveraging Shifted Delta Cepstra (SDC) features derived from Perceptually Enhanced Zero-Time Windowed Cepstral Coefficients (PE-ZTWCCs), we demonstrated the effectiveness of a shallow Time Delay Neural Network (TDNN) classifier. Our results show that optimizing the temporal contexts of SDC with TDNN significantly boosts classification performance, underscoring the importance of feature engineering in neural networks, especially for smaller datasets. These findings, along with the novel datasets, open promising avenues for future research in speech pathology and disfluency classification.
\section{Acknowledgement}
The authors would like to thank The Indian Stammer Association (TISA) for their invaluable collaboration in data collection and for their insightful discussions.

\newpage
\bibliographystyle{IEEEtran}
\bibliography{references}

\begin{thebibliography}{10}
\providecommand{\url}[1]{#1}
\csname url@samestyle\endcsname
\providecommand{\newblock}{\relax}
\providecommand{\bibinfo}[2]{#2}
\providecommand{\BIBentrySTDinterwordspacing}{\spaceskip=0pt\relax}
\providecommand{\BIBentryALTinterwordstretchfactor}{4}
\providecommand{\BIBentryALTinterwordspacing}{\spaceskip=\fontdimen2\font plus
\BIBentryALTinterwordstretchfactor\fontdimen3\font minus \fontdimen4\font\relax}
\providecommand{\BIBforeignlanguage}[2]{{%
\expandafter\ifx\csname l@#1\endcsname\relax
\typeout{** WARNING: IEEEtran.bst: No hyphenation pattern has been}%
\typeout{** loaded for the language `#1'. Using the pattern for}%
\typeout{** the default language instead.}%
\else
\language=\csname l@#1\endcsname
\fi
#2}}
\providecommand{\BIBdecl}{\relax}
\BIBdecl

\bibitem{lickley2017disfluency}
R.~Lickley, ``Disfluency in typical and stuttered speech,'' \emph{Book series Studi AISV}, vol.~3, pp. 373--387, 2017.

\bibitem{yairi2013epidemiology}
E.~Yairi and N.~Ambrose, ``Epidemiology of stuttering: 21st century advances,'' \emph{Journal of fluency disorders}, vol.~38, no.~2, pp. 66--87, 2013.

\bibitem{apple}
\BIBentryALTinterwordspacing
``Improved speech recognition for people who stutter,'' 2023. [Online]. Available: \url{https://machinelearning.apple.com/research/speech-recognition}
\BIBentrySTDinterwordspacing

\bibitem{boey2009awareness}
R.~A. Boey, P.~H. Van~de Heyning, F.~L. Wuyts, L.~Heylen, R.~Stoop, and M.~S. De~Bodt, ``Awareness and reactions of young stuttering children aged 2--7 years old towards their speech disfluency,'' \emph{Journal of communication disorders}, vol.~42, no.~5, pp. 334--346, 2009.

\bibitem{starkweather1982stuttering}
C.~W. Starkweather, ``Stuttering and laryngeal behavior: A review,'' \emph{Asha Monographs}, no.~21, pp. 1--45, 1982.

\bibitem{kuniszyk1996comparison}
W.~Kuniszyk-J{\'o}{\'z}kowiak, ``A comparison of speech envelopes of stutterers and nonstutterers,'' \emph{The Journal of the Acoustical Society of America}, vol. 100, no.~2, pp. 1105--1110, 1996.

\bibitem{wingate2002foundations}
M.~E. Wingate and P.~Howell, ``Foundations of stuttering,'' \emph{The Journal of the Acoustical Society of America}, vol. 112, no.~4, pp. 1229--1231, 2002.

\bibitem{shriberg2005spontaneous}
E.~Shriberg, ``Spontaneous speech: how people really talk and why engineers should care.'' in \emph{INTERSPEECH}.\hskip 1em plus 0.5em minus 0.4em\relax Citeseer, 2005, pp. 1781--1784.

\bibitem{lea2021sep}
C.~Lea, V.~Mitra, A.~Joshi, S.~Kajarekar, and J.~P. Bigham, ``Sep-28k: A dataset for stuttering event detection from podcasts with people who stutter,'' in \emph{Proc. ICASSP}.\hskip 1em plus 0.5em minus 0.4em\relax IEEE, 2021, pp. 6798--6802.

\bibitem{9528931}
T.~Kourkounakis, A.~Hajavi, and A.~Etemad, ``Fluentnet: End-to-end detection of stuttered speech disfluencies with deep learning,'' \emph{IEEE/ACM Transactions on Audio, Speech, and Language Processing}, vol.~29, pp. 2986--2999, 2021.

\bibitem{mathews2018language}
S.~M. Mathews, ``Language skills and secondary education in india,'' \emph{Economic and Political Weekly}, vol.~53, no.~15, pp. 20--22, 2018.

\bibitem{garg2021towards}
S.~Garg, U.~Mehrotra, G.~Krishna, and A.~K. Vuppala, ``Towards a database for detection of multiple speech disfluencies in indian english,'' in \emph{2021 National Conference on Communications (NCC)}.\hskip 1em plus 0.5em minus 0.4em\relax IEEE, 2021, pp. 1--6.

\bibitem{bayerl_KSoFKasselState_2022}
S.~Bayerl, A.~Wolff~von Gudenberg, F.~H{\"o}nig, E.~Noeth, and K.~Riedhammer, ``Ksof: {The Kassel State of Fluency Dataset -- A Therapy Centered Dataset of Stuttering},'' in \emph{Proceedings of the Language Resources and Evaluation Conference}.\hskip 1em plus 0.5em minus 0.4em\relax Marseille, France: European Language Resources Association, Jun. 2022, pp. 1780--1787.

\bibitem{russell2005magnitude}
M.~Russell, M.~Corley, and R.~J. Lickley, ``Magnitude estimation of disfluency by stutterers and nonstutterers,'' in \emph{Phonological encoding and monitoring in normal and pathological speech}.\hskip 1em plus 0.5em minus 0.4em\relax Psychology Press, 2005, pp. 260--272.

\bibitem{arbisi2005comparison}
T.~Arbisi-Kelm and S.-A. Jun, ``A comparison of disfluency patterns in normal and stuttered speech,'' in \emph{Disfluency in Spontaneous Speech}, 2005.

\bibitem{kadiri2018breathy}
S.~R. Kadiri and B.~Yegnanarayana, ``Breathy to tense voice discrimination using zero-time windowing cepstral coefficients (ztwccs).'' in \emph{Interspeech}, 2018, pp. 232--236.

\bibitem{simha2023enhancing}
N.~V.~R. Simha, M.~S. Ganesh, and V.~A. Kumar, ``Enhancing stutter detection in speech using zero time windowing cepstral coefficients and phase information,'' in \emph{International Conference on Speech and Computer}.\hskip 1em plus 0.5em minus 0.4em\relax Springer, 2023, pp. 130--141.

\bibitem{bayya2013spectro}
Y.~Bayya and D.~N. Gowda, ``Spectro-temporal analysis of speech signals using zero-time windowing and group delay function,'' \emph{Speech Communication}, vol.~55, no.~6, pp. 782--795, 2013.

\bibitem{makur2001warped}
A.~Makur and S.~K. Mitra, ``Warped discrete-fourier transform: Theory and applications,'' \emph{IEEE Transactions on Circuits and Systems I: Fundamental Theory and Applications}, vol.~48, no.~9, pp. 1086--1093, 2001.

\bibitem{hermansky1990perceptual}
H.~Hermansky, ``Perceptual linear predictive (plp) analysis of speech,'' \emph{the Journal of the Acoustical Society of America}, vol.~87, no.~4, pp. 1738--1752, 1990.

\bibitem{kim2016power}
C.~Kim and R.~M. Stern, ``Power-normalized cepstral coefficients (pncc) for robust speech recognition,'' \emph{IEEE/ACM Transactions on audio, speech, and language processing}, vol.~24, no.~7, pp. 1315--1329, 2016.

\bibitem{alluri2017detection}
K.~R. Alluri, S.~Achanta, S.~R. Kadiri, S.~V. Gangashetty, and A.~K. Vuppala, ``Detection of replay attacks using single frequency filtering cepstral coefficients.'' in \emph{Interspeech}, 2017, pp. 2596--2600.

\bibitem{peddinti2015time}
V.~Peddinti, D.~Povey, and S.~Khudanpur, ``A time delay neural network architecture for efficient modeling of long temporal contexts.'' in \emph{Interspeech}, 2015, pp. 3214--3218.

\bibitem{torres2002approaches}
P.~A. Torres-Carrasquillo, E.~Singer, M.~A. Kohler, R.~J. Greene, D.~A. Reynolds, and J.~R. Deller~Jr, ``Approaches to language identification using gaussian mixture models and shifted delta cepstral features.'' in \emph{Interspeech}.\hskip 1em plus 0.5em minus 0.4em\relax Citeseer, 2002, pp. 89--92.

\bibitem{shriberg1994preliminaries}
E.~Shriberg, ``Preliminaries to a theory of speech disfluency,'' \emph{PhD Diss, Univ. of California}, 1994.

\bibitem{freeman1979phonation}
F.~J. Freeman, ``Phonation in stuttering: A review of current research,'' \emph{Journal of Fluency Disorders}, vol.~4, no.~1, pp. 79--89, 1979.

\end{thebibliography}
\end{document}